\documentclass[a4paper,11pt]{article}
\usepackage{pos}
\usepackage{comment}
\def\nn{\nonumber}
\def\be{\begin{equation}}             \def\ee{\end{equation}}
\def\ba#1{\begin{array}{#1}}          \def\ea{\end{array}}
\def\bea{\begin{eqnarray} }           \def\eea{\end{eqnarray} }
\def\beann{\begin{eqnarray*} }        \def\eeann{\end{eqnarray*} }
\def\beal{\begin{eqalign}}            \def\eeal{\end{eqalign}}
             
\def\bsubeq{\begin{subequations}}     \def\esubeq{\end{subequations}}
\def\bitem{\begin{itemize}}           \def\eitem{\end{itemize}}
                    
\def\pa{\partial}

\def\a{\alpha}
\def\b{\beta}

\def\d{\delta}

\def\g{\gamma}

\def\l{\lambda}
\def\m{\mu}
\def\n{\nu}
\def\o{\omega}
\def\r{\rho}                    
\def\s{\sigma}                  
\def\t{\tau}

\def\ds{\stackrel{\star}{,}}
\newcommand{\dwedge}{\curlywedge}
\newcommand{\sfR}{\mathsf{R}}

\def\G{\Gamma}

\title{Noncommutative fields in Reissner–Nordström black hole background}

\author[a]{Milorad Be\v zani\'c}
\author[a]{Marija Dimitrijevi\'c \'Ciri\'c}
\author*[a]{Nikola Konjik}
\author[a]{Biljana Nikoli\'c}
\author[b]{Andjelo Samsarov}

\affiliation[a]{University of Belgrade, Faculty of Physics\\
Studentski trg 12, Belgrade, Serbia}

\affiliation[b]{Institute Rudjer Bo\v skovi\'c\\
Bijeni\v cka cesta 54, Zagreb, Croatia}

\emailAdd{milorad.bezanic@ff.bg.ac.rs}
\emailAdd{dmarija@ipb.ac.rs}
\emailAdd{konjik@ipb.ac.rs}
\emailAdd{biljana@ipb.ac.rs}
\emailAdd{asamsarov@irb.hr}

\abstract{In this short paper we discuss dynamics of noncommutative (NC) matter fields in the Reissner–Nordström (RN) black hole background. After reviewing the propagation of charged NC scalar and spinor fields, we derive the equation governing the propagation of NC electromagnetic (EM) perturbation in the RN background. The propagation of NC scalar and spinor perturbation have a dual description in terms of the propagation of commutative fields in the effective/dual metric. Finally, we turn to the gravitational perturbations. We present equations of motion for the NC gravitational field obtained in two different models: $SO(2,3)_\star$ NC gravity and braided NC gravity. Typically for NC gravity models, the first nontrivial corrections are quadratic in the NC parameter. The obtained NC gravity equations are the starting point to discuss the propagation of NC gravitational perturbations and the validity of the dual description in terms of the effective metric.}

\FullConference{
}

\begin{document}
\maketitle

\section{Introduction}

Since the discovery of the black hole (BH) solution of Einstein's equation by Schwarzschild \cite{SchwBH} more than one century ago, BH physics has been a very active and intriguing field of research. Although BHs are fully described just by their mass $M$, electric and/or magnetic charge $Q/P$, and angular momentum $J$, their properties are far from being fully understood. In parallel to the theoretical research, there have been groundbreaking observations of BHs and their mergers recently: LIGO and Virgo collaborations detected gravitational waves arising
from the mergers of binary black holes \cite{GWIntro1}, while the Event Horizon Telescope (EHT) produced images of the supermassive BH
in the center of the M87 galaxy \cite{BHIntro1}.

Much of the insight regarding gravitational radiation and properties of BHs may be gained by studying BH
perturbations. Once perturbed, a BH responds by going through a ringdown phase, during which it starts emitting gravitational waves. The most important part of the ringdown phase is a long period of damped oscillations, dominated by the quasinormal modes (QNMs). Besides carrying the intrinsic information about the BH itself, QNMs may also
carry information about the properties of the underlying spacetime structure. In particular, if the underlying spacetime
structure is deformed in such a way that it departs from the usual notion of a smooth manifold, then this deviation should leave its trace in the QNMs’ spectrum too.

To study QNM spectra in the nonommutative (NC) geometry setting, it is necessary to first derive equations of motion for NC  fields in a fixed gravitational background. We will focus on the twist formalism approach to NC geometry and to the Reissner–Nordström (RN) black hole gravitational background. RN BH is a solution of the Einstein-Maxwell equations in the presence of a point electric charge $Q$ and magnetic charge $P$ in the origin $r=0$. Although astrophysical BHs are expected to be nearly neutral due to charge neutralization by surrounding plasma, charged BHs can serve as valuable laboratories to probe the nature of horizons, singularities and some aspects of quantum gravity. They are also important in holography and AdS/CFT, where charged black holes in anti-de Sitter space correspond to finite-temperature states in strongly coupled quantum field theories with a chemical potential \cite{BHIntro2}. Recently, there have been suggestions that a charged BH can arise in the presence of dark matter halos \cite{RNIntro1}.

In this short paper we discuss dynamics of NC matter fields in the RN black hole background. The NC deformation is introduced via the angular twist \cite{rnqnm}. We will review the dynamics of NC scalar and spinor fields in the RN black hole background and then discuss the dynamics of the NC electromagnetic field and the NC gravitational field. In the next section, we briefly review the angular twist deformation and introduce the notation. Then we discuss dynamics of the NC matter fields: scalar, spinor and vector fields in Section 3. Finally, in Section 4 we present equations of motion for the NC gravitational field obtained in two different models. We postpone a detailed discussion on the propagation of the NC gravitational perturbations for future work.

\section{Angular twist deformation}

The NC deformation is introduced via the twist operator of the form 
\begin{equation}
\mathcal{F} = e^{-\frac{i}{2}\theta ^{\alpha\beta}X_\alpha\otimes X_\beta} = e^{-\frac{ia}{2} (\partial_t\otimes\partial_\varphi - \partial_\varphi\otimes\partial_t)}\, ,
\label{AngTwist0Phi}
\end{equation} 
with the small NC parameter $a$. Vector fields $X_{1}=\partial _t$,
$X_{2}= \partial_\varphi$ are commuting vector fields,
$[X_{1},X_{2}]=0$, therefore the twist (\ref{AngTwist0Phi}) is an Abelian twist \cite{PL09}. It is dubbed as "angular twist" becuase the vector field $X_2 =
\partial_\varphi$ is a generator of rotations around $z$-axis. 
The twist (\ref{AngTwist0Phi}) defines the $\star$-product of functions,
\begin{align}
f\star g =&  \mu \{ e^{\frac{ia}{2} (\partial_t\otimes \partial_\varphi - \partial_\varphi\otimes
\partial_t)}
f\otimes g \}\nn\\
=& fg + \frac{ia}{2}
(\partial_t f(\partial_\varphi g) - \partial_t g(\partial_\varphi f)) + 
\mathcal{O}(a^2) .\label{fStarg0Phi}
\end{align}
The algebra of differential forms is deformed in the similar way \cite{rnqnm}. Noncommutativity of $\star$-products is controled by the triangular $R$-matrix ${\cal R} =
{\cal F}_{21}{\cal F}^{-1}$, with   ${\cal R}_{21} = \sfR_{\alpha}\otimes\sfR^{\alpha} = {\cal R}^{-1}$. In particular:
\begin{align}
f\star g =&  \sfR_{\alpha}(g)\star \sfR^{\alpha}(f)  \, ,  \nn\\
\omega_1\wedge_\star\omega_2 =& (-1)^{p_1p_2}\, \sfR_{\alpha}(\omega_2)\wedge_\star \sfR^{\alpha}(\omega_1)\, .\nn
\end{align}

Let us fix the background geometry to that of the RN black hole with the electric charge $Q\neq 0$ and no magnetic charge. The metric tensor is then given by
\begin{equation}
{\rm d}s^2 = (1-\frac{2MG}{r}+\frac{Q^2G}{r^2}){\rm d}t^2 - \frac{{\rm
d}r^2}{1-\frac{2MG}{r}+\frac{Q^2G}{r^2}} - r^2({\rm d}\theta^2 + \sin^2\theta{\rm d}\varphi^2), \label{dsRN}
\end{equation}
where $M$ and $Q$ are the mass and charge of the RN black hole, respectively. The vector fields $X_1$ and $X_2$ are two Killing vectors for this metric, thus the twist
(\ref{AngTwist0Phi}) does not act on the RN metric. In this way we ensure that the background geometry remains intact by the deformation. 

\section{NC matter fields in the RN black hole background}

The NC actions for the charged scalar field $\hat{\phi}$, the charged spinor field $\hat{\psi}$ and the massless vector field $\hat{A}_\mu$ are given by \cite{rnqnm, Dimitrijevic-Ciric:2022vvl}
\begin{align}
S^\star_\phi =& \int {\rm d}^4x \, \sqrt{-g}\star\Big( g^{\mu\nu}\star D_{\mu}\hat{\phi}^+ \star
D_{\nu}\hat{\phi} - \mu^2\hat{\phi}^+ \star\hat{\phi}\Big) , \label{SPhi}\\
S^\star_\psi =& \int {\rm d}^4x\, \sqrt{-g} \star\bar{\hat{\Psi}} \star\Big( i\gamma^\mu \big( \partial_{\mu} \hat{\Psi} - i\hat{\omega}_\mu \star \hat{\Psi} -iq\hat{A}_\mu\star \hat{\Psi} \big) -m\hat{\Psi}\Big) ,\label{SPsi}\\
S^\star_A =& -\frac{1}{4q^2} \int {\rm d}^4 x\,\sqrt{-g}\star g^{\alpha\beta}\star g^{\mu\nu}\star
\hat{F}_{\alpha\mu}\star \hat{F}_{\beta\nu} .\label{SA}
\end{align}
The NC scalar field  $\hat{\phi}$ has the mass $\mu$ and charge $q$ and  transforms in the fundamental representation of NC $U(1)_\star$. The NC spinor field  $\hat{\psi}$ has the mass $m$ and charge $q$ and it also transforms in the fundamental representation of NC $U(1)_\star$. The NC $U(1)$  gauge field $\hat{A}_\mu$ is introduced in the model  through a minimal coupling and the NC field-strength tensor is defined as
\begin{equation}
\hat{F}_{\m\nu} = \partial_\mu \hat{A}_\nu - \partial_\nu \hat{A}_\mu - i [A_\mu \ds A_\nu]. \label{NCF}
\end{equation}
The covariant derivatives are defined as
\begin{align}
D_\mu\hat{\phi} =& \partial_\mu\hat{\phi} - i \hat{A}_\mu\star \hat{\phi}, \nonumber\\
D_\mu\hat{\Psi} =& \partial_\mu\hat{\Psi} - i \hat{\omega}_\mu\star \hat{\Psi} - i \hat{A}_\mu\star \hat{\Psi}, \nn
\end{align}
with the spin connection $\hat{\omega}_\mu = \frac{i}{2}  \hat{\omega}_{\mu}^{~~ab} \Sigma_{ab}$ and $\Sigma^{ab}$ are the generators of the local Lorentz transforamtions.

The actions  (\ref{SPhi})-(\ref{SA}) are invariant under the infinitesimal $U(1)_\star$
gauge transformations:
\begin{align}
\delta^\star \hat{\phi} =& i\hat{\Lambda} \star \hat{\phi}, \quad \delta^\star \hat{\Psi} = i\hat{\Lambda} \star \hat{\Psi}\, ,\nn\\
\delta^\star \hat{A}_\mu =& \partial_\mu\hat{\Lambda} + i[\hat{\Lambda} \ds \hat{A}_\mu],
\label{NCGaugeTransf}\\
\delta_\star \hat{F}_{\mu\nu} =& i[\hat{\Lambda} \ds \hat{F}_{\mu\nu}],\nn\\
\delta_\star \hat{\omega}_{\mu\nu} =& 0,\quad \delta^\star g_{\mu\nu} = 0.\nn
\end{align}
where $\hat{\Lambda}$ is the NC gauge parameter. The last two transformations in (\ref{NCGaugeTransf}) make clear that the models studied here are semiclassical. By this we mean that only scalar, spinor and gauge field are subject to a NC deformation, while gravitational field remains  unaffected. A more general situation dealing with a NC deformation of gravitational field becomes  increasingly more involved. For more details see \cite{PL09}, \cite{PLSWGeneral}, \cite{SO23} and references therein.

In order to expand the actions (\ref{SPhi})-(\ref{SA}) up to first order in the deforamtion parameter $a$, we systematically expand all $\star$-products and use the Seiberg-Witten (SW) map. SW map enables to express NC variables as functions of the corresponding commutative variables \cite{SW}. In this
way, the problem of charge quantization in $U(1)_\star$ gauge theory does not exist. In the case of
NC Yang-Mills gauge theories, SW map guarantees that the number of degrees of freedom in the NC
theory is the same as in the corresponding commutative theory. That is, no new degrees of freedom
are introduced.

Using the SW map NC fields can be expressed as function of corresponding commutative fields
and can be expanded in orders of the deformation parameter $a$. Expansions for an arbitrary Abelian
twist deformation are known to all orders \cite{PLSWGeneral}. Applying these results to the twist
(\ref{AngTwist0Phi}), 
expansions of fields up to first
order in the deformation parameter $a$ follow. They are given by\footnote{In the spherical coordinate system $x^\mu = (t,r,\theta,\varphi)$ the twist (\ref{AngTwist0Phi}) reduces to the Moyal-like twist
\begin{equation}
{\cal F} = e^{-\frac{i}{2}\theta ^{\alpha\beta}\partial_\alpha\otimes \partial_\beta} \, .\nn    
\end{equation}
The matrix $\theta^{\alpha\beta}$ is related to the NC paramter $a$ as:
\begin{equation}
\begin{pmatrix}
0&0&0& a\\
0&0&0& 0\\
0&0&0& 0\\
-a&0&0& 0\\
\end{pmatrix}  \, .\label{Theta-a}  
\end{equation}}:
\begin{align}
\hat{\phi} =& \phi -\frac{1}{4}\theta^{\rho\sigma}A_\rho(\partial_\sigma\phi + D_\sigma
\phi), \label{HatPhi}\\
\hat{\Psi} =& \Psi -\frac{1}{4}\theta^{\rho\sigma}A_\rho(\partial_\sigma\Psi + D_\sigma
\Psi), \label{HatPsi}\\
\hat{A}_\mu =& A_\mu -\frac{1}{2}\theta^{\rho\sigma}A_\rho(\partial_\sigma A_{\mu} +
F_{\sigma\mu}), \label{HatA}  \\
\hat{F}_{\mu\nu} =& F_{\mu\nu} - \frac{1}{2}\theta^{\rho\sigma}A_{\rho}(\partial_\sigma F_{\mu\nu}
+ D_\sigma F_{\mu\nu})+\theta^{\rho\sigma}F_{\rho\mu}F_{\sigma\nu}. \label{HatFmunu}
\end{align}

In the following we discuss equations of motion for each of the three matter fields.

\subsection{Scalar fields}

Using the SW map solutions and expanding the $\star$-products in (\ref{SPhi}) we
find the NC scalar field action up to first order in the deformation parameter $a$. It is given by
\begin{align}
S^\star_\phi =& \int
{\rm d}^4x\sqrt{-g}\,
\Big( g^{\mu\nu}D_\mu\phi^+D_\nu\phi -\mu^2\phi^+\phi \label{SExp}\\
&+ \frac{\theta^{\alpha\beta}}{2}g^{\mu\nu}\big( -\frac{1}{2}D_\mu\phi^+F_{\alpha\beta}
D_\nu\phi
+(D_\mu\phi^+)F_{\alpha\nu}D_\beta\phi + (D_\beta\phi^+)F_{\alpha\mu}D_\nu\phi\big) \Big)
.\nn 
\end{align}
Varying the action \eqref{SExp} with respect to $\phi^\dagger$ one obtains an equation of motion for $\phi$
\begin{eqnarray}\label{eom}
  && g^{\mu\nu}  \Big[D_{\mu}D_{\nu}\phi-\Gamma^{\lambda}_{\mu\nu}D_{\lambda}\phi -\frac{1}{4}\Theta^{\alpha\beta}\big( D_\mu(F_{\alpha\beta}D_\nu\phi) \nonumber  \\   
&-& \Gamma^{\lambda}_{\mu\nu}F_{\alpha\beta}D_\lambda\phi-2D_\mu(F_{\alpha\nu}D_\beta\phi)+2\Gamma^{\lambda}_{\mu\nu}F_{\alpha\lambda}D_\beta\phi-2D_\beta(F_{\alpha\mu}D_\nu\phi) \big)\Big]=0.
\end{eqnarray}
In  (\ref{SExp}) the coupling constant $q$ is absorbed into a definition of $A_\mu$, so that $A_\mu \rightarrow qA_\mu$.

Since the RN background also fixes the background electromagnetic setting, the corresponding $U(1)$ gauge field $A_\mu$ and the corresponding field strength tensor $F_{\mu\nu}$ are given by
\begin{equation}
A_0 = -\frac{qQ}{r}, \qquad  F_{r0} = \frac{qQ}{r^2}\, .   \label{A0}
\end{equation}
Taking into account (\ref{A0}), one obtains the equation of motion in the form
\begin{align}
&
\Big( \frac{1}{f}\partial^2_t -\Delta + (1-f)\partial_r^2 
+\frac{2MG}{r^2}\partial_r + 2iqQ\frac{1}{rf}\partial_t -\frac{q^2Q^2}{r^2f}\Big)\phi
\nonumber\\
& +\frac{aqQ}{r^3}
\Big( (\frac{MG}{r}-\frac{GQ^2}{r^2})\partial_\varphi
+ rf\partial_r\partial_\varphi \Big) \phi =0, \label{EomPhiExp1}
\end{align}
where $\Delta$ is the usual Laplace operator. 

It is shown in \cite{Dimitrijevic-Ciric:2022vvl} that equation (\ref{EomPhiExp1}) can be rewritten as
\begin{equation} 
{\Box_{g'}}  \phi \equiv  g'^{\mu \nu} \big( \nabla'_{\mu} - i A_{\mu}  \big )  \big( \nabla'_{\nu} - i A_{\nu}  \big)   \phi = \frac{1}{\sqrt{-g'}} (\partial_{\mu} -iA_{\mu}) \bigg( \sqrt{-g'} ~ g'^{\mu \nu} \big( \partial_{\nu} - i A_{\nu} ) \bigg)  \phi = 0 \, , \nn
\end{equation}
with the dual metric $g^{\prime}_{\mu\nu}$.
\begin{equation}  \label{1stordereffmetrica}
g^{\prime}_{\mu\nu}=\begin{pmatrix}
-f&0&0&0\\
0&\frac{1}{f}&0&\frac{aqQ}{2}\sin^2\theta\\
0&0&r^2&0\\
0&\frac{aqQ}{2}\sin^2\theta&0&r^2 \sin^2\theta\\
\end{pmatrix}+\mathcal{O}(a^2).
\end{equation}
The additional off-diagonal terms in (\ref{1stordereffmetrica}) are induced purely by noncommutative nature of spacetime. The equation of motion (\ref{EomPhiExp1}) can thus be understood as the equation of motion governing behaviour of a charged commutative scalar field (having the same charge $q$ as its NC counterpart), propagating in a modified RN geometry
\begin{equation} \label{NCdsRN}   
  {\rm d}s^2 = \Big(1-\frac{2MG}{r}+\frac{Q^2G}{r^2} \Big) {\rm d}t^2 - \frac{{\rm
d}r^2}{1-\frac{2MG}{r}+\frac{Q^2G}{r^2}} - aqQ \sin^2 \theta {\rm d} r {\rm d} \phi - r^2({\rm d}\theta^2 + \sin^2\theta{\rm d}\phi^2 ).
\end{equation}

\subsection{Spinor fields}

The expanded NC spinor action (\ref{SPsi}) up to first order in $a$ is given by
\begin{align}
S^\star_\Psi =& \int {\rm d} ^4x ~ \sqrt{-g} \Big\{ \bar{\Psi} \Big( i\gamma^\mu D_\mu \Psi -m\Psi\Big) \nn\\
& +\frac{1}{2}\theta^{\alpha\beta}\Big( -iF_{\mu\alpha}\bar{\Psi}\gamma^\mu D^{\mbox{\tiny{U(1)}}}_\beta\Psi
-\frac{i}{2}\bar{\Psi}\gamma^\mu \omega_\mu F_{\alpha\beta}\Psi 
-\frac{1}{2}F_{\alpha\beta}\bar{\Psi} \big(i\gamma^\mu D^{\mbox{\tiny{U(1)}}}_\mu \Psi -m\Psi\big) \Big) \Big\}.\nn
\end{align}
Remembering that $F_{\alpha\beta} = \partial_\alpha A_\beta - \partial_\beta A_\alpha$ and
choosing the electromagnetic potential to be that of the RN black hole,
the only non-zero component of $F_{\alpha\beta}$ is $F_{r0} = \frac{qQ}{r^2}$. This leads to a simplified NC action
\begin{equation}
S^\star_\Psi = \int {\rm d} ^4x ~ |e| \Big( \bar{\Psi} \Big( i\gamma^\mu D_\mu \Psi -m\Psi\Big) -\frac{i}{2}\theta^{\alpha\beta}\bar{\Psi} F_{\mu\alpha}\gamma^\mu (\partial_\beta\Psi -iA_\beta\Psi) \Big)\label{SNCExpanded}
\end{equation}
and the corresponding equation of motion for the spinor $\Psi$
\begin{equation}
i\gamma^\mu \Big(\partial_\mu \Psi -i \omega_\mu\Psi - iA_\mu\Psi\Big) -m\Psi - \frac{ia}{2}F_{rt}\gamma^r\partial_\phi \Psi = 0 .
\end{equation}
Inserting the explicit expressions for $F_{rt}$ and $\gamma^{r} = e_a^{~~r} \gamma^{a},$ this equation reduces to
\begin{equation}     \label{EoMPsiNC} 
   i\gamma^\mu \Big(\partial_\mu \Psi -i \omega_\mu\Psi - iA_\mu\Psi\Big) -m\Psi -
    \frac{ia}{2}  \frac{qQ}{r^2} \sqrt{f} \gamma^1 \partial_{\phi} \Psi = 0 .
\end{equation}
This equation may serve for studying the spinor quasinormal modes in the RN background \cite{QNMSpinors}.

As we mentioned in the scalar case, in the spinor case we can also establish the  equivalence between commutative
and noncommutative theories at the level of the equation of motion. Equation \eqref{EoMPsiNC} can be rewritten as
\begin{equation}
\Big( i \gamma^a e_a^{'~~\mu} ( \partial_\mu -i \omega'_{\mu} -i A_{\mu} ) - m \Big)\Psi =0\, .\nn
\end{equation}
The dual vierbein $e^{'a}_{~~\mu}$ and the dual spin connection $\omega'_{\mu}$ can be related with the dual metric (\ref{1stordereffmetrica}) \cite{Dimitrijevic-Ciric:2022vvl}. In particular
\begin{equation} \label{metrictetrad}
  e^{'a}_{~~\mu} =
\left( \begin{array}{ccccc}
  \sqrt{f} & 0  & 0 & 0  \\
   0   & \frac{1}{\sqrt{f}} & 0 & 0  \\ 
   0  & 0 & r &  0 \\
 0  & \frac{aqQ }{2r} \sin \theta & 0 &  r \sin \theta \\
\end{array} \right) \, .
\end{equation}

It is very intriguing that for both NC scalar and NC spinor fields the duality between the NC fields in the commutative background and the commutative fields in the dual NC background holds with the same dual geometry described by (\ref{NCdsRN}). In our recent work \cite{DimitrijevicCiric:2024ibc} we have studied properties of the dual metric (\ref{NCdsRN}) in more details.

\subsection{Vector fields}

Using the SW-map solutions and expanding the $\star$-products in (\ref{SA}) we
find the NC EM field action up to first order in the deformation parameter $a$. It is given by
\begin{equation}
S_A = \int
{\rm d}^4x\sqrt{-g}\,
\Big( -\frac{1}{4q^2}g^{\mu\rho}g^{\nu\sigma}F_{\mu\nu}F_{\rho\sigma}
+\frac{1}{8q^2}g^{\mu\rho}g^{\nu\sigma}\theta^{\alpha\beta}(F_{\alpha\beta}F_{\mu\nu}F_{\rho\sigma}
-4F_{\mu\alpha}F_{\nu\beta}F_{\rho\sigma}) \Big).\label{SAexp} 
\end{equation}

Now we vary the action (\ref{SAexp}) with respect to $A_\lambda$ to calculate the equations of motion: 
\begin{align}
&\partial_\mu F^{\mu\lambda} + \Gamma^{\rho}_{\mu \rho} F^{\mu\lambda}
+\theta^{\alpha\beta}
\Big( -\frac{1}{2}\big( \partial_\mu (F_{\alpha\beta}F^{\mu\lambda}) + \Gamma^{\rho}_{\mu \rho}
F_{\alpha\beta}F^{\mu\lambda} \big) \nn\\
& + \partial_\mu (F_{\alpha}^{\ \mu}F_\beta^{\ \lambda}) +
\Gamma^{\rho}_{\mu \rho}F_{\alpha}^{\ \mu}F_\beta^{\ \lambda}
-\partial_\alpha(F_{\beta\mu}F^{\mu\lambda}) \Big) \nn\\
& +\theta^{\alpha\lambda}
\Big( \frac{1}{2}\big( \partial_\mu (F_{\beta\nu}F^{\mu\nu})
+\Gamma^\rho_{\mu\rho}F_{\beta\nu}F^{\mu\nu}\big) -\frac{1}{4}\partial_\alpha(F_{\mu\nu}F^{\mu\nu})
\Big) = 0 .\label{EoMA}
\end{align}

Taking into account that EM field has two components, one from RN background and  other from external dinamical EM field, we can write $F_{\mu\nu}=F^{RN}_{\mu\nu}+\epsilon\mathcal{F}_{\mu\nu}$, where $\mathcal{F}_{\mu\nu}$ is external EM field and $\epsilon$ is small parameter. As expected, the zeroth order in $\epsilon$ is trivial because the RN EM field $F^{RN}_{\mu\nu}$ is the solution of the Einstein-Maxwell equations with the only non zero component $F^{RN}_{r0} = \frac{qQ}{r^2}$. The first order in $\epsilon$ is given by
\begin{align}\label{eomA}
   &\partial_r\mathcal{F}^{r0}+\partial_\varphi\mathcal{F}^{\varphi 0}+\partial_\theta\mathcal{F}^{\theta 0}+\frac{2}{r}\mathcal{F}^{r0}+\cot{\theta}\mathcal{F}^{\theta 0}+\frac{aqQ}{r^2}(-\frac{1}{2}\partial_r\mathcal{F}_{0\varphi}+\frac{1}{2}\partial_0\mathcal{F}_{\varphi r}-\frac{2}{r}\mathcal{F}_{\varphi 0}) =0,\nonumber\\
   &\partial_0\mathcal{F}^{0r}+\partial_\varphi\mathcal{F}^{\varphi r}+\partial_\theta\mathcal{F}^{\theta r}+\cot{\theta}\mathcal{F}^{\theta r}+\frac{aqQ}{r^2}(\partial_0\mathcal{F}_{\varphi 0}-\partial_\theta\mathcal{F}_{\varphi \theta}-\cot{\theta}\mathcal{F}_{\varphi\theta}) =0,\nonumber\\
   &\partial_0\mathcal{F}^{0\theta}+\partial_\varphi\mathcal{F}^{\varphi \theta}+\partial_r\mathcal{F}^{r\theta}+\frac{2}{r}\mathcal{F}^{r\theta}+\frac{aqQ}{r^2}(\partial_r\mathcal{F}_{\varphi \theta}-\partial_\varphi\mathcal{F}_{r \theta}) =0,\\
   &\partial_0\mathcal{F}^{0\varphi}+\partial_r \mathcal{F}^{r\varphi}+\partial_\theta\mathcal{F}^{\theta \varphi}+\cot{\theta}\mathcal{F}^{\theta \varphi}+\frac{2}{r}\mathcal{F}^{r\varphi}+\frac{aqQ}{r^2}(\frac{1}{2}\partial_\varphi\mathcal{F}_{r\varphi}-\frac{1}{2}\partial_\theta\mathcal{F}_{\theta r}+\frac{1}{2}\cot{\theta}\mathcal{F}_{\theta r}) =0.\nonumber
\end{align}

In order to analyze the dual description as in the case of NC scalar and NC spinor fields, we need to rewrite equations \eqref{eomA} in the following form
\begin{equation}\label{dualA}
\partial_\mu F^{\mu\lambda} + \Gamma'^{\rho}_{\mu \rho} F^{\mu\lambda}=0.
\end{equation}
Here we introduced the Christoffel symbols for dual metric $\Gamma'^{\rho}_{\mu \rho}$. It is obvious that equations \eqref{eomA} cannot be rewritten in the  form \eqref{dualA} because terms proportional to $a\,\partial \mathcal{F}$ cannot be absorbed in Christoffel symbols for the dual metric $\Gamma'$. The failure of the dual description for the NC EM field might be a consequence of the nonlinearity of the SW map for the gauge field $\hat{A}_\mu$ (\ref{HatA}) which results in terms cubic in $F_{\mu\nu}$ in the NC action (\ref{SAexp}). After the variation, these terms produce terms of the form $a\,\partial (F \,F)$ that spoil the dual description. On the other hand, the SW maps for the NC scalar field $\hat{\phi}$ (\ref{HatPhi}) and the NC spinor field $\hat{\Psi}$ (\ref{HatPsi}) are linear in the commutative fields $\phi$ and $\Psi$ and therefore the dual picture holds.

\section{NC gravitational field}

There are different approaches to the construction of a NC deformation of gravity, see \cite{NCGrReviews} and references there in. Here we will focus on two different models: $SO(2,3)_\star$ NC gravity \cite{SO23} based on the $\star$-NC gauge symmetry and the braided NC gravity \cite{BraidedGr} based on the braided NC gauge symmetry.

Gravity models based on the $SO(2,3)$ gauge symmetry appeared in the 1970ies \cite{SO23old}. The original motivation for those models was to study properties of supergravity. The $SO(2,3)$ algebra reduces via the Inon\" u-Wigner contraction to the usual Poincar\' e algebra in $4D$ spacetime. This
is used to obtain General Relativity from (A)dS gauge theory.

\vspace{2mm}
The NC generalization of the action proposed in \cite{SO23old} is given by
\begin{equation}
S^\star_{g} = c_1S^\star_{1g} + c_2S^\star_{2g} + c_3S^\star_{3g} \, ,\label{NCSO23Action}
\end{equation}
with
\begin{align}
S^\star_{1g} =& \frac{il}{64\pi G_N}{\rm Tr} \int
\hat{F} \wedge_\star \hat{F}\star \hat{\phi}\, ,\nn\\ 
S^\star_{2g} =& \frac{1}{64 \pi G_{N}l}{\rm Tr} \int \hat F \wedge_\star \hat{D\phi} \wedge_\star{\hat D\phi} \star \hat{\phi} + c.c. \, , \nn \\
S^\star_{3g} =& -\frac{i}{128 \pi G_{N}l}{\rm Tr}\int \hat{D\phi}\wedge_\star \hat{D\phi} \wedge_\star{\hat D\phi}\wedge_\star \hat{D\phi} \star \hat{\phi} \,
.\nn 
\end{align}
We introduced here the $SO(2,3)_\star$ field strength tensor $\hat{F}$ and the scalar field $\hat{\phi}$ transforming in the adjoint representation of $SO(2,3)_\star$
\begin{align}
\hat{F} =& \frac{1}{2}\hat{F}_{\mu\nu}^{AB}M_{AB} \star {\rm d}x^\mu \wedge_\star {\rm d}x^\nu =
\frac{1}{2}\hat{F}_{\mu\nu}^{AB}M_{AB} {\rm d}x^\mu \wedge {\rm d}x^\nu ,\quad \widehat{\delta}_{\widehat{\rho}}\widehat{F}_{\mu\nu}=i[\widehat{\rho}\ds\widehat{F}_{\mu\nu}] ; \label{FSO23}\\
\hat{\phi} =& \hat{\phi}^A \Gamma_A, \quad \widehat{\delta}_{\widehat{\rho}}\hat{\phi}=i[\widehat{\rho}\ds\hat{\phi}]\nn
\end{align}
with the $SO(2,3)$ algebra generators, $M_{AB}=-M_{BA}$ ($A,B=0,1,2,3,5$) and the $5$D gamma-matrices $\Gamma^A$. We choose to work in the spherical coordinate system that is adapted to the twist (\ref{AngTwist0Phi}), therefore the $\star$-products between the basis one forms and basis one forms and functions reduce to the ordinary commutative multiplication. 

To obtain a NC $D=4$ gravity action, we proceed as follows: Firstly, we expand $\star$-products and insert the SW map solutions for the NC fields $\hat{F}_{\mu\nu}$ and $\hat{\phi}$ into the action (\ref{NCSO23Action}). As expected, the NC correction linear in the NC parameter $\theta^{\mu\nu}$ disappears and the first nontrivial NC correction is quadratic in $\theta^{\mu\nu}$. The SW expansion guarantees that the NC action is invariant unded the commutative $SO(2,3)$ gauge symmetry. Next, we break the $SO(2,3)$ symmetry to the $SO(1,3)$ by fixing the value of the background field to $\phi=l\gamma_{5}$. 

The obtained result is very cumbersome \cite{SO23}, and we will not write the full expression here. However, one can still analyze the model in different regimes of parameters. If we are interested in the low-energy sector of the theory, we should keep only terms that have at most two derivatives on vierbeins. Therefore, we include only terms linear in curvature and linear and quadratic in torsion (we assumed that the spin connection and the first-order derivatives of vierbeins are of the same order). The equations of motion are obtained by varying the action over vierbein and spin connection, independently. If we consider only the class of NC solutions with classically vanishing torsion $T_{\mu\nu}^{a(0)}=0$ (where $T_{\mu\nu}^{a(0)}=0$ is the zeroth order in $\theta^{\mu\nu}$ expansion of the full NC torsion), in the low-energy limit, equations of motion can be written as 
\begin{equation} 
R^{\n\m}-\frac12g^{\m\n}R+\frac{3}{l^2}(1+c_2+2c_3) g^{\m\n}=\tau^{\m\n} 
,\label{EoM-metric-0}
\end{equation}
with
\begin{align}
\tau^{\m\n} =&-\frac{\theta^{\a\b}\theta^{\g\d}}{16l^4}\Big(
(3-3c_2+24c_3)(2R^\n_{\ \b\g\d}\d_\a^\m+R_{\a\b\g\d}g^{\m\n})\nn\\
& +(4-18c_2-44c_3)(2R^\n_{\ 
\g\b\d}\d_\a^\m+R_{\a\g\b\d}g^{\m\n})\nn\\
& -(6+22c_2+36c_3)(g^{\m\n}
g_{\b\d}R_{\a\g}+2\d_\b^\m \d_\d^\n R_{\a\g}+g_{\b\d}\d_\g^\m R_{\a}^{\ 
\n}-g_{\b\d}R_{\a\ \g}^{\ \n\ \m})\nn\\ 
& +(20+28c_2-8c_3)(2\d_\b^\m
\G^{\n}_{\g\a}\G^{\r}_{\d\r}
-g^{\m\n}g_{\s\tau}\G^{\s}_{\d\b}\G^{\tau}_{\g\a}-\frac12g^{\m\n}R_{\a\b\g\d}
)\nn\\
&+(7-25c_2-28c_3)(g^{\m\n}g_{\tau\d}\G^\r_{\a\g}\G^\tau_{\b\r}+
g^{\m\n}g_{\rho\tau}\G^\r_{\a\g}\G^\tau_{\b\d}
-\d^\m_\g g^{\rho\nu}g_{\tau\d}\G^\s_{\a\s}\G^\tau_{\b\r}\nn\\
&-\d_\g^\m
\G^\s_{\a\s}\G^\n_{\b\d}+\d^\m_\g 
g^{\s\nu}g_{\tau\d}\G^\r_{\a\s}\G^\tau_{\b\r}+\d^\m_\g 
g_{\rho\tau}g^{\s\n}\G^\r_{\a\s}\G^\tau_{\b\d}- 
g^{\rho\nu}g_{\tau\d}\G^\m_{\a\g}\G^\tau_{\b\r}\nn\\
&-\G^\m_{\a\g}\G^\n_{\b\d}-\d^\m_\d\G^\s_{\b\s}\G^\n_{\a\g}-\d^\n_\d 
\G^\s_{\b\s}\G^\m_{\a\g}\nn\\
&+\d^\m_\d 
\G^\s_{\a\g}\G^\n_{\b\s}+\d^\n_\d 
\G^\s_{\a\g}\G^\m_{\b\s}+\frac12\d_\d^\n R_{\a\b\ \g}^{\ \ \m}+\frac12\d_\d^\m 
R_{\a\b\ \g}^{\ \ \n})\nn\\
& +(2+4c_2+8c_3)(g^{\m\n}g_{\b\d}(2\pa_\a\G^\r_{\g\r}+\G^\s_{\g\r}\G^\r_{\a\s}+ 
\G^\s_{\g\s}\G^\r_{\a\r})+(\d_\b^\m\d_\d^\n+\d_\b^\n\d_\d^\m)(\pa_\a\G^\r_{\g\r}
+\G^\s_{\g\r}\G^\r_{\a\s})
\nn\\
& +2g^{\m\n}g_{\s\b }
\G^\r_{\g\r}\G^\s_{\s\d}-2
g^{\s\n}g_{\d\b }
\G^\r_{\a\r}\G^\m_{\s\g}
-2\pa_\g\G^{\m}_{\a\s}g^{\s\n}g_{\b\d}\nn\\
&
-2g_{\b\rho}g^{\tau\n}\G^\m_{
\g\tau}
\G^{\r}_{\a\d}+g_{\b\tau}g^{\m\n}\pa_\g\G^\tau_{
\a\d}\nn\\
&+ g^{\m\n}
\G^{\tau}_{\a\d}(g_{\tau\s}\G^{\s}_{\g\b}+g_{\b\s}\G^{\s}_{\g\tau})+\frac{
6+28c_2+56c_3}{l^2}(g^{\m\n}g_{\a\g}g_{\b\d}+4g_{\b\d}
\d_\g^\n\d_\a^\m)\Big)\ .\label{EoM-metric}
\end{align}
Expanding the NC metric as $g_{\mu\nu} = g_{\mu\nu}^{(0)} + h_{\mu\nu}$, equation (\ref{EoM-metric}) can be used to study gravitational perturbations in the RN background. Developing further ideas from \cite{Herceg:2024vwc}, we plan to study gravitational perturbations in our future work.

Recently, another approach to NC gauge theories was introduced in \cite{BraidedGr}. Unlike the $\star$-gauge transformations (\ref{NCGaugeTransf}), braided NC gauge transformations close the Lie algebra. Therefore, no new degrees of freedom appear and the SW map is not necessary. The NC braided gravity action in $D=4$ dimensions is given by \cite{BraidedGr}
\begin{align}
S^{\star}_{(e,\omega)} =& \int_M\,{\rm Tr}\Big(\frac12\,e\dwedge_\star e\dwedge_\star R^\star + \frac\Lambda4 \, e\dwedge_\star e\dwedge_\star e\dwedge_\star e\Big) \label{BraidedGrAction} \\ 
& -\frac1{24} \, \int_M\,{\rm Tr}\Big(\omega\dwedge_\star\big(2\,e\dwedge_\star T_{\mbox{\tiny L}}^\star - 2\,T_{\mbox{\tiny R}}^\star\dwedge_\star e + {\rm D}_{\star{\mbox{\tiny L}}}^\omega(e\dwedge_\star e) + {\rm D}_{\star{\mbox{\tiny R}}}^\omega(e\dwedge_\star e)\big)\Big) \ .\nn    
\end{align}
The curvature $R^\star$ and the torsion $T_{\mbox{\tiny L,R}}$ are defined as
\begin{align}
R^\star =& {\rm d} \omega + \frac{1}{2}[\omega, \omega]_\star, \quad [\omega, \omega]_\star = \omega\wedge_\star \omega + \sfR_{\alpha} \omega \wedge_\star \sfR^{\alpha}\omega \,; \nn\\
T_{\mbox{\tiny L}} =& {\rm d} e + \omega\wedge_\star e, \>\> T_{\mbox{\tiny R}} = {\rm d} e + e\wedge_\star \omega .\nn
\end{align}
The action (\ref{BraidedGrAction}) is invariant under the braided diffeomorphism symmetry and under the braided gauge $SO(1,3)$ symmetry. In particular, the curvature and the torsion are covariant under the braided gauge $SO(1,3)$ transformations
\begin{align}
\delta_\rho R^\star =& [\rho, R^\star]_\star =  \rho\star R^\star + \sfR_{\alpha} R^\star \star \sfR^{\alpha}\rho \,; \nn\\ 
\delta_\rho T_{\mbox{\tiny L}} =& [\rho, T_{\mbox{\tiny L}}]_\star =  \rho\star T_{\mbox{\tiny L}} + \sfR_{\alpha} T_{\mbox{\tiny L}} \star \sfR^{\alpha}\rho \, , \nn
\end{align}
with the NC gauge parameter $\rho = \frac{1}{2}\rho^{ab}\Sigma_{ab}$.

The equations of motion are obtained by varying the action (\ref{BraidedGrAction}) with respect to the NC spin connection and the NC vierbein and they are exact in the NC parameter $\theta^{\mu\nu}$ \cite{BraidedGr}. In order to compare these equations with the equations following form the $SO(2,3)_\star$ model, in particular with (\ref{EoM-metric}), we now expand the exact equations of motion up to the 2nd order in the NC parameter $\theta^{\mu\nu}$. 

The expanded equation of motion for the NC vierbein $e^a_\m$ is given by
\begin{equation}
{R_d}^{\s}-\frac12e^\s_d R={\t_d}^{\s} \, , \label{BraidedEoMe}
\end{equation}
with
\begin{align}
{\t_d}^\s =& \frac{1}{8}\theta^{\a\b}\theta^{\g\d} \big[ \frac{1}{2}(\pa_\a \pa_\g e^a_\m) (\pa_\b \pa_\d \pa_\n  \o^{bc}_\r) + \frac{1}{2} e_\m^a (\pa_\a \pa_\g \o^{bf}_\n) (\pa_\b \pa_\d {\o_{\r f}}^c) \nn\\ 
&+ \frac{2}{3} (\pa_\a e_\m^a) (\pa_\b \pa_\d \o^{bf}_\n) (\pa_\g {\o_{\r f}}^c) + (\pa_\a \pa_\g e^a_\m) (\pa_\b \pa_\d  \o^{bf}_\n) {\o_{\r f}}^c \nn\\
&+ \frac{1}{3} (\pa_\a \pa_\g e^a_\m) (\pa_\b  \o^{bf}_\n) (\pa_\d {\o_{\r f}}^c)\big] 
\frac{1}{e}
\varepsilon^{\mu\nu\rho\sigma}
\varepsilon_{abcd} \ .
\end{align}

The expanded equation of motion for the spin connection is given by
\begin{equation}
    {T_{ac}}^a e_d^\d - {T_{ad}}^a e_c^\d + {T_{cd}}^\d ={S_{cd}}^\d\ , \label{BraidedEoMomega}
\end{equation}
with
\begin{align}
{S_{cd}}^\d =& \frac{1}{8} \theta^{\a\g}\theta^{\b\d} \Bigg[ (\pa_\a \pa_\g e_{\m}^a)(\pa_\b \pa_\d \pa_\r e^b_\n) + 
(\pa_\a \pa_\g \o_\m^{ac})(\pa_\b \pa_\d e_{\n c}) e_\r^b 
+ (\pa_\a \pa_\g \o_\m^{ac}) e_{\n c} (\pa_\b \pa_\d e_\r^b) \nn \\
& + \o_\m^{ac} (\pa_\a \pa_\g e_{\n c}) (\pa_\b \pa_\d e_\r^b) 
+ \frac{2}{3} (\pa_\a \pa_\g \o_\m^{ac}) (\pa_\b 
e_{\n c}) (\pa_\d e_\r^b) \nn \\
& +  \frac{2}{3} (\pa_\b \o_\m^{ac}) (\pa_\a \pa_\g
e_{\n c}) (\pa_\d e_\r^b) + \frac{2}{3} (\pa_\b \o_\m^{ac}) (\pa_\d
 e_{\n c}) (\pa_\a \pa_\g e_\r^b)
\Bigg]\frac{1}{e}\varepsilon^{\a\b\g\d} \varepsilon_{abcd}\ .
\end{align}

In order to study gravitational perturbations around the background metric in the braided formalism, we have to solve equation (\ref{BraidedEoMomega}) and express the NC spin connection in terms of the NC vierbein. This solution should then be inserted into equations (\ref{BraidedEoMe}) to obtain an analogue of (\ref{EoM-metric}). This analysis we postpone for our future work.

\section{Conclusions and outlook}

In this short paper we have first reviewed our earlier work on the charged NC scalar and spinor fields in the RN black hole background (\ref{dsRN}). We showed that there is a dual interpretation of our results: They can be understood as propagation of commutative charged scalar and spinor fields in the effective/dual metric (\ref{NCdsRN}) with the corrections generated by the NC deformation. The equations (\ref{EomPhiExp1}) and (\ref{EoMPsiNC}) are used in \cite{rnqnm} and \cite{QNMSpinors} to analyse the NC QNM spectra of the RN black hole.

Next, the propagation of the NC EM perturbation is discussed in some detail. We have calculated the equation of motion in the RN background up to first order in the NC parameter $a$. The dual description fails in this case, due to the nonlinearity of the SW map for the gauge field $\hat{A}_\mu$.

Finally, we presented two different approaches to the NC gravity: $SO(2,3)_\star$ gravity and the braided NC gravity. For both models we calculated the expanded equations of motion. The first nontrivial correction in both cases is quadratic in the NC parameter $\theta^{\mu\nu}$. Starting from these equations, in our future work we plan to discuss gravitational perturbations in the RN black hole background. 

\paragraph{Acknowledgments.}
We thank the organizers of the Corfu Summer Institute
2024 for the stimulating meeting and the opportunity to present the
results of our work. This research was supported by the
Croatian Science Foundation Project IP-2020-02-9614 Search for Quantum spacetime in Black Hole QNM spectrum
and Gamma Ray Bursts.
The work of {\sc M.B., M.D.C. N.K. and B.N.} is supported by Project
451-03-136/2025-03/200162 of the Serbian Ministry of Science,
Technological Development and Innovation. 


\end{document}